 \documentclass[aps,prb,reprint,showpacs,longbibliography]{revtex4-1}

  \usepackage{graphicx}
  \usepackage{dcolumn}
  \usepackage{longtable}
  \usepackage{bm}

\newcommand{\s}{$\;\,$}
\newcommand{\RK}{$R_{mt}K_{max}$}

\begin{document}

\title{Improved generalized gradient approximation for positron states in solids}

\author{Jan Kuriplach}
\email{Jan.Kuriplach@mff.cuni.cz}
\affiliation{
Department of Low Temperature Physics, Faculty of Mathematics and Physics, Charles University, V Hole\v{s}ovi\v{c}k\'ach 2, CZ-180\,00 Prague, Czech Republic}

\author{Bernardo Barbiellini}
\email{bba@neu.edu}
\affiliation{
Department of Physics, Northeastern University, Boston, Massachusetts 02115, USA}


  \begin{abstract}
Several first-principles calculations of positron-annihilation characteristics in solids have added gradient corrections to the local-density approximation within the theory by Arponen and Pajanne [Ann. Phys. (N.Y.) {\bf 121}, 343 (1979)] since
this theory systematically overestimates the annihilation rates.
As a further remedy we propose to use gradient corrections for other local density approximation schemes based on perturbed hypernetted-chain and on Quantum Monte Carlo results.
Our calculations for various metals and semiconductors show that the proposed schemes generally improve the positron lifetimes when they are confronted with experiment.
We also compare the resulting positron affinities in solids with data from slow-positron measurements.
  \end{abstract}


\maketitle

\section{Introduction}
The Density Functional Theory (DFT) solves for the
electronic structure of a condensed matter
system in its ground state so that the electron density
$\rho^-$ is the basic quantity.\cite{review_dft}
The DFT can be generalized to positron-electron
systems by including the positron density $\rho^+$ as well
and it is then called two-component DFT.\cite{tdft,review_tdft}
As a consequence of the Hohenberg-Kohn theorem \cite{review_dft}
the ground-state expectation value of any
operator $\hat{o}$ is a functional of the electron and positron
densities denoted by $O[\rho^-,\rho^+]$.
It can be shown \cite{bauer} that if the Hamiltonian $\hat{{\cal H}}$
is augmented by the operator
through a scalar coupling $\eta$,
\begin{equation}
\hat{{\cal H}}=\hat{{\cal H}}(\eta{=}0)+\eta \hat{o}\,,
\end{equation}
and the exchange-correlation (XC) energy $E_{xc}[\rho^-,\rho^+](\eta)$
is computed for small values of the field $\eta$,
then the correction to the expectation value calculated using
the Kohn-Sham single-determinant wave function is the derivative
of the XC energy with respect to the field $\eta$:
\begin{equation}
O[\rho^-,\rho^+]=O_0[\rho^-,\rho^+]+
\left.\frac{d}{d\eta} E_{xc}[\rho^-,\rho^+](\eta) \right|_{\eta=0},
\end{equation}
where $O_0[\rho^-,\rho^+]$ is the expectation
value of $\hat{o}$
for a system of noninteracting fermions moving
in the effective field provided by the Kohn-Sham
formalism,\cite{review_dft}
$E_{xc}$ is the XC energy functional.
This general expression for $O[\rho^-,\rho^+]$
generalizes the Lam-Platzman theorem \cite{bauer}
and provides a formal scheme
to extract positron annihilation
characteristics from the
two-component DFT.\cite{mrs99}

The Local Density Approximation (LDA) is the simplest
implementation
of the DFT \cite{review_tdft,review_dft} and it provides
an explicit formula for $E_{xc}$.
The generalized gradient approximation (GGA)
gives a systematic improvement for first-principles
electronic calculations with respect to the LDA.\cite{perdew,gga1,gga2,alatalo2,gga3}
In the case of a positron impurity embedded in an electron
gas, the LDA based on the theory by Arponen and Pajanne\cite{AP}
underestimates systematically the positron lifetime
and the positron affinity while for the GGA,\cite{gga1,gga2} the agreement with
the experiment in real materials improves since
the annihilation rate contains density gradient corrections
which reduces electron-positron correlation effects.
This reduction is important in the regions of core and semicore electrons
and largest interstitial spaces in semiconductors.

Nevertheless, the GGA correction to the LDA based on the
Arponen and Pajanne scheme does not completely cancel the
overestimation of the annihilation rate in Al.\cite{gga1}
Moreover other limitations of this GGA correction
\cite{mitroy,campillo,ggaboro}
have revealed the need to develop new semilocal functionals.
Therefore the goal of the present paper is to
study how the GGA for positron states in solids could be
improved by replacing the Arponen and Pajanne LDA scheme
with LDA parametrizations based on perturbed hypernetted-chain\cite{phnc} (PHNC)
 and on recent Quantum Monte Carlo\cite{qmc} (QMC) results.
Preliminary results of Boro\'nski\cite{ggaboro} have suggested
that the GGA based on PHNC is
a right step towards an improved GGA scheme for positron states
in materials.  Here, we report state-of-the-art self-consistent all electron
calculation without any shape approximation to the charge density
or potential to check this interesting hypothesis.

The calculations of positron annihilation characteristics are not only important to test various approaches to the electron-positron correlation problem but are also useful for other applications. 
For instance, positron annihilation is widely used in condensed matter physics and in materials science to study Fermi surfaces\cite{West} and open volume defects\cite{Hautojarvi} in the bulk and at near surface regions of materials. 
Accurate calculations of positron characteristics are therefore needed in order to reliably extract physically sound results from experiments.\cite{Tuomisto}

An outline of this paper is as follows.
Sec. \ref{LDAGGA} deals with the basic principles of LDA and GGA for positrons.
In Sec. \ref{Methods}, we present the details of the electronic structure and 
positron annihilation characteristics calculations. 
The results of the calculations are presented and compared with experimental results
in Sec. \ref{RD}, and the conclusions are given in Sec. \ref{End}.

\section{LDA and GGA for positrons} \label{LDAGGA}

\begin{figure}[h]
  \begin{center}
  \includegraphics[width=8cm,height=6cm]{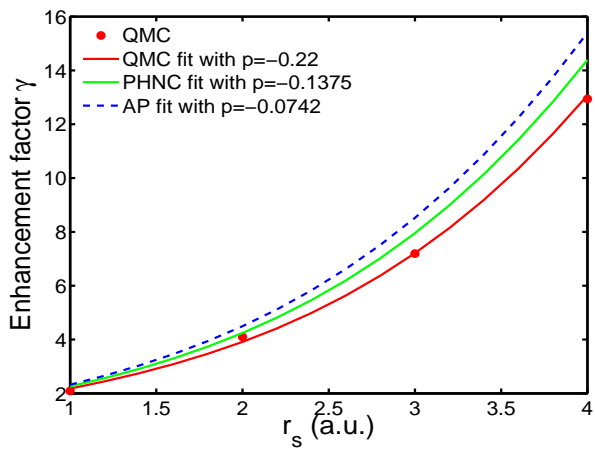}
  \end{center}
  \caption{(Color online) Enhancement factor $\gamma$ as a function of $r_s$.
  The AP theory yields the higher enhancement
  while PHNC and QMC give lower values.
  All the enhancements factors are fitted with
  the formula given by Eq.~(\ref{eq:gamma}).
  The only fitting parameter is the coefficient
  $p$ of the square $r_s$ term.}
  \label{fig1}
  \end{figure}

The shape of the screening cloud around a positron in a given material is similar to that of a positronium (Ps) atom and it determines the positron lifetime  through the positron-electron contact density, which is enhanced by a factor $\gamma$ with respect to the unperturbed electron density. In the LDA, $\gamma$ is treated as a function of the local electron density. If the positron is considered as an impurity, a useful LDA parametrization of $\gamma$ as a function of the electron gas parameter $r_s$ reads as
\begin{equation}
\gamma= 1 + 1.23\,r_s + p\,r_s^2 + r_s^3/3~.
\label{eq:gamma}
\end{equation}
In Eq.~(\ref{eq:gamma}) the factor $p$ in front
of the square term is the only fitting parameter
while the first two terms are fixed to reproduce
the high-density
RPA limit and the last term the low-density
Ps atom limit.\cite{bba_review}
The value $p=-0.0742$ parametrizes
the results by Arponen and Pajanne\cite{gga1} (AP)
while the value $p=-0.1375$ fits the
perturbed hypernetted chain approximation\cite{phnc} (PHNC).
The recent Quantum Monte Carlo data can be fitted
with $p=-0.22$ as shown in Fig.~\ref{fig1}.
Moreover, near $r_s=2$ the Boro\'nski-Nieminen method,\cite{tdft} the Jarlborg-Singh model,\cite{js,bgj} the Sterne and Kaiser parametrization,\cite{sterne} and the QMC enhancement predict almost the same result $\gamma \approx 4$. Increasing $p$ within the LDA gives better agreement with the experiment than the Arponen and Pajanne $p=-0.0742$ used in the same type of approximation.\cite{ggaboro}

However, quite generally, the LDA underestimates the positron lifetime of solids.
In fact one expects that the strong electric field due to the inhomogeneity
suppresses the electron-positron correlations in
the same way as the Stark effect decreases the electron-positron density
at zero distance for the Ps atom. In the
GGA, the effects of the nonuniform electron density are described in terms of
the ratio between the local length scale  of the density variations
$|\nabla \ln \rho^-|$ and the local Thomas-Fermi screening length $1/q_{TF}$.
The lowest order gradient correction to the
LDA correlation hole density
is proportional to the parameter
$\epsilon=(|\nabla \ln \rho^-|/q_{TF})^2$.
This parameter is taken to describe also the reduction of
the screening cloud close to the positron.
For the homogeneous electron gas $\epsilon= 0$,
whereas in the case of rapid density variations
$\epsilon$ approaches infinity.
At the former limit the LDA result for the induced screening charge is valid
and the latter limit should lead to the independent particle model
with vanishing enhancement.
In order to interpolate between these limits,
one uses for the enhancement factor the form
\begin{equation}
\gamma_{GGA} - 1= (\gamma_{LDA}-1) \exp(-\alpha \epsilon)~.
\end{equation}
Where $\alpha$ is set so that the calculated and experimental lifetimes
agree as well as possible for a large number of different types of solids.
The corresponding electron-positron correlation potential scales as \cite{gga1}
\begin{equation}
V_{GGA}^{ep} = V_{LDA}^{ep} \exp(-\alpha \epsilon/3)~.
\end{equation}
For the Arponen and Pajanne LDA $\alpha$ is $0.22$ (Ref. \onlinecite{gga1})
while for the PHNC the value must be renormalized to $0.10$ (Ref. \onlinecite{ggaboro}).
We will see that $\alpha=0.05$ in the case of LDA based on QMC results.

Further in the text, we shall use the following abbreviations for various LDA and GGA positron approaches examined:
BN for the Boro\'nski and Nieminen approach,\cite{tdft}
GC for the gradient correction with the AP theory and $\alpha=0.22$ (after Ref. \onlinecite{gga1}),
SL for the Stachowiak and Lach PHNC theory,\cite{phnc}
SG for the gradient correction with the SL approach and $\alpha=0.10$ (after Ref. \onlinecite{ggaboro}),
DB for the Drummond {\em et al.} QMC theory,\cite{qmc} and
DG for the gradient correction with the DB approach and $\alpha=0.05$. 

\section{Computational methods}  \label{Methods}

The electronic structure calculations for selected crystalline solids, metals and semiconductors,
were carried out using the WIEN2k code.\cite{wien2k}
This code implements the augmented plane wave plus local orbital (APW+lo) method,\cite{APWlo}
which is considered to be one of the most accurate methods to calculate electronic structure of solids, and is
based on the the linearized augmented plane wave (LAPW) method.\cite{LAPW}
The recommended option to use the mixed APW+lo/LAPW basis set was chosen in the present work.
The WIEN2k program also performs full-potential calculations,
which impose no shape restrictions for the electron density and potential.
The LDA electron XC based on QMC simulations
by Ceperley and Alder\cite{eCAXC} and parametrized by Perdew and Wang\cite{eCAPWXC}
was employed to perform the electronic structure calculations.
The effects of the electron gradient corrections were also tested with the GGA
functional by Perdew, Burke, and Ernzerhof\cite{ePBEXC} in several solids.

In order to obtain the positron wave function ($\psi^+$) and energy ($E^+$),
a computer code was developed based on a finite difference
method\cite{ATSUP1,ATSUP2} to solve
the positron Schr\"odinger equation (in Hartree atomic units)
\begin{equation}
\left[-\frac12\nabla_{\bm{r}}^2 - V^c(\bm{r}) + V^{ep}(\bm{r}) \right] \psi^+(\bm{r}) =
E^+ \psi^+(\bm{r})
\label{pSe}
\end{equation}
for which the electron Coulomb potential ($V^c$) and total electron density ($\rho^-$)
needed to determine $V^{ep}$ are taken from self-consistent WIEN2k electronic structure calculations.
In particular, the positron potential and wave function are calculated on a regular 3D mesh,
which covers an appropriately chosen crystallographic unit cell of the studied solid.
Other computational details are given in the Appendix (Sec. \ref{APC}).

The positron lifetime ($\tau$) is calculated via the positron annihilation
rate ($\lambda$) according to the formula\cite{tdft}
\begin{eqnarray}
\frac{1}{\tau} = \lambda &=& \pi r_0^2 c \int \textrm{d}\bm{r}\,
\rho^+\!(\bm{r})\, \rho^-\!(\bm{r})\, \gamma[\rho^-\!(\bm{r}),\epsilon(\bm{r})] \nonumber\\
&=& \int \textrm{d}\bm{r}\, \lambda_i(\bm{r})
\label{eq:lambda}
\end{eqnarray}
where $r_0$ and $c$ are the classical electron radius and speed of light, respectively.
The spatial integration proceeds over the unit cell.
The $\lambda_i$ symbol denotes the integrand of the annihilation rate.
The positron density is obtained simply as $\rho^+ = |\psi^+|^2$
from the properly normalized positron wave function
since only the positron ground state is considered.

In addition to the positron lifetime, the positron affinity $A^+$
is also an important bulk property of materials.
The positron affinity can be determined\cite{review_tdft} as
\begin{equation}
A^+=\mu^- +\mu^+ = -(\phi^- + \phi^+)\,,
\label{PAF}
\end{equation}
where $\mu^-$ and $\mu^+$ are the electron and positron chemical
potentials, respectively.
Within the context of present calculations
$\mu^+$ can be identified with the ground state positron energy $E^+$.
In the case of a semiconductor, $\mu^-$ is taken from the position
of the top of the valence band.
The comparison of measured and calculated positron affinity values for different materials is a good test for the electron-positron correlation potential $V^{ep}$.
From the experimental viewpoint, the positron affinity can be found by measuring the electron ($\phi^-$) and positron ($\phi^+$) work functions, as also documented in Eq. (\ref{PAF}).
The positron affinity (work function) is usually measured by positron re-emission spectroscopy.\cite{schultz,mills} 
Alternatively, the positron affinity can be obtained using the positronium formation potential.\cite{schultz,Si100pWF}
A method based on the examination of positron-induced secondary electron spectra was suggested mainly for cases with positive positron work function in Ref. \onlinecite{PNPAF}, but to our knowledge it has not been used in practice so far.
Experimental determination of the positron affinity requires the knowledge of the electron work function.
Such measurements are often based on the contact potential difference method (e.g. Kelvin probe) which has recently been doubted to be really related to this difference for semiconductors.\cite{CPD}
This might lead to the revision of some experimental results related to positron affinity determination.

Qualitatively speaking, positron properties depend on the
average electron density and thereby on the unit cell volume.
For instance, from Eq. (\ref{eq:lambda}) one can deduce that the positron lifetime
will increase with the decreasing average electron density.
In the case of Al, the numerical test shows that increasing the lattice constant by 1\% results in an increase of the positron lifetime by 2.2\%.
It is therefore desirable to have precise crystal structure parameters of investigated materials.
For the purpose of the present study, we have considered room temperature lattice constants,\cite{LCs} except stated otherwise.

In order to assess the precision of our calculations, we performed the check of various numerical parameters of the WIEN2k code. 
The details of such checks are described in the Appendix (Sec. \ref{AWP}) and the most important parameter was found to be the basis set cutoff.
Thus, we present in the next section well converged positron lifetimes and affinities with a numerical precision of the order of 0.1 ps and 0.01 eV, respectively.
The statistical accuracy of experimental counterparts is typically around 1 ps and 0.1 eV.
Therefore, a reliable comparison of our calculations with available experimental data is warranted.
Furthermore, our way of calculation, which
avoids non self-consistent atomic superpositions\cite{ATSUP1,ATSUP2}
or shape approximations\cite{bdj} to the charge density or potential,
allows us to assess with great precision the effects of various LDA and GGA
correlation schemes on the positron characteristics without the perturbation
from any numerical approximations.

Regarding defect studies, we focused on positron trapped at ideal monovacancy in Al, Si and Cu.
Therefore, the ions neighboring the monovacancy are not allowed to relax from their ideal lattice positions.
This approximation is expected to be a good one in metals but in semiconductors the  lattice relaxation
may depend strongly on the charge state of the vacancy.\cite{gga2,Makkonen06}
To calculate the electronic structure of defects, we employed a supercell
approach placing the vacancy in the center of supercells.
The supercells containing 107, 215, and 107 atoms for Al, Si and Cu, respectively
were constructed from the perfect fcc (Al, Cu) or diamond (Si)
with 3 $\times$ 3 $\times$ 3 cubic unit cells.
Such supercell sizes are adequate to obtain vacancy properties related
to electronic structure like the vacancy formation energy.
On the other hand, there exist problems\cite{Korhonen96} with the positron in vacancy supercell calculations.
One hundred atom or similar supercells are usually too small for the accurate positron wave function determination using the periodic boundary condition which is natural for supercell calculations.
The point is that the wave function is overly delocalized, which results in too short positron lifetime and too small positron binding energy compared to isolated vacancy.
This is the effect of finite supercell size.
To approach isolated vacancy behavior would therefore require much larger supercells, which might not be computationally feasible.

Korhonen \emph{et al.} suggest\cite{Korhonen96} to perform a $\bm{k}$-space integration for low lying positron states to correct the positron wave function behavior in smaller supercells.
Here, we use another procedure to obtain the correct positron wave function in supercell calculations.
Following previous positron computational studies of defects (see e.g. Ref. \onlinecite{Brauer09}) we add atoms in the form of regular lattice at the sides of our supercells.
The electron density and the electron Coulomb potential for this added lattice are taken from separate WIEN2k (bulk) calculation and are continuously joined to those of the supercell.
Further details of this approach are explained in the Appendix (Sec. \ref{ADC}).
The determination of positron properties then proceeds in the way similar to the bulk calculations: The positron lifetime is computed using Eq. (\ref{eq:lambda}) and the positron binding energy is obtained as a difference between the bulk and supercell positron energies.

\section{Results and discussions}  \label{RD}

\subsection{Bulk positron lifetime and affinity}

The positron lifetime results are presented in Table \ref{tabL} while Table \ref{tabA} gives the affinities, both quantities being calculated according to various approaches to electron-positron correlations summarized above.
The tables contain results for selected elements including metals (alkalies, transition metals, {\it sp}-metals, lanthanides), semiconductors (group IV) and a solid inert gas (Ne).
As for compounds, most of them are semiconductors (group IV, II-VI, IV-VI) and one intermetallics.
In our opinion, this list of elements and compounds is well suited to test our approach to electron-positron correlations based on the gradient correction.

\begin{table*}[htb]
\caption{Positron lifetimes (in ps) calculated according to various approaches:
BN = Boro\'nski and Nieminen,\cite{tdft}
AP = Arponen and Pajanne,\cite{AP}
GC = gradient correction with AP ($\alpha=0.22$),
SL = Stachowiak and Lach,\cite{phnc}
SG = gradient correction with SL ($\alpha=0.10$),
DB = Drummond {\em et al.},\cite{qmc}
DG = gradient correction with DB ($\alpha=0.05$).
The 10th column gives experimental values taken from the reference specified in the last column or Refs. \onlinecite{seeger,perexp03,sourceEldrup} (see the text).
In the comment column FM and NM stand for ferromagnetic and non-magnetic states, respectively. 
The information about temperature is related to the positron measurement temperature and to structural data used for calculations.
The symbol $\overline{\tau}$ means that only average lifetimes are available for $\alpha$-Sn, $\alpha$-Ce and $\gamma$-Ce.
Marks in columns GC and DG evaluate the closennes of theory and experiment: A -- deviation $<$5 ps, B -- deviation $<$10 ps and C -- deviation $>$10 ps (see the text for details).}
\begin{ruledtabular}
\begin{tabular}[t]{@{\ }l l r r r r r r r c l c}
System & Structure & BN\s& AP\s& GC\s\s\s& SL\s& SG\s& DB\s& DG\s\s\s& Experiment & Comment & Ref. \\
\hline\\[-2mm]
\multicolumn{11}{c}{Elements} \\
Li  & bcc       & 299.5 & 259.4 & 283.4 B& 277.2 & 295.0 & 303.5 & 315.8 C & 291 & & \onlinecite{expalkali} \\
C   & diamond   &  92.8 &  86.7 & 102.3 A&  90.5 &  98.1 &  94.6 &  98.8 A & 98+ & & \onlinecite{Dannefaer00} \\
Ne  & fcc       & 234.2 & 218.2 & 531.5 C& 229.4 & 370.8 & 242.4 & 312.0 C & 430$\pm$30 &
20\,K & \onlinecite{liu1963} \\
Na  & bcc       & 328.1 & 290.7 & 338.6 A& 309.9 & 341.6 & 342.8 & 364.2 C & 338 & & \onlinecite{expalkali} \\
Al  & fcc       & 165.0 & 146.1 & 154.8 B& 154.6 & 160.1 & 161.3 & 164.5 A & 160+ \\
Si  & diamond   & 211.0 & 185.7 & 221.7 A& 197.2 & 214.3 & 208.3 & 217.2 A & 216+ \\
Fe  & bcc       & 101.1 &  93.6 & 109.9 A&  97.9 & 106.5 & 102.1 & 106.5 A & 105+ & FM \\
Cu  & fcc       & 106.4 &  98.4 & 120.3 B& 102.9 & 114.4 & 107.3 & 113.2 A & 110+ \\
Zn  & hcp       & 137.1 & 124.3 & 149.8 A& 130.7 & 144.2 & 136.4 & 143.6 B & 145+ \\
Ge  & diamond   & 214.2 & 189.8 & 240.8 C& 201.2 & 225.2 & 212.8 & 225.2 A & 219+ \\
Nb  & bcc       & 121.4 & 110.5 & 123.4 A& 116.1 & 122.7 & 120.9 & 124.3 A & 120+ \\
Ag  & fcc       & 123.1 & 112.7 & 140.4 B& 118.2 & 131.9 & 123.4 & 130.5 B & 130+ \\
Sn  & diamond, $\alpha$-Sn& 257.4 & 227.8 & 296.3 B& 242.1 & 273.8 & 258.3 & 274.9 C& 289 & $\overline{\tau}$ & \onlinecite{SnLifetime} \\
Sn  & $\beta$-Sn\,(A5)& 188.7 & 166.9 & 197.6 A& 176.8 & 192.0 & 186.2 & 194.2 A& 196+ \\
Ce  & fcc, $\alpha$-Ce& 167.9 & 149.1 & 169.7 C& 157.7 & 168.0 & 165.2 & 170.6 C& 233 & NM, 80\,K, $\overline{\tau}$ & \onlinecite{CePAS} \\
Ce  & fcc, $\gamma$-Ce& 197.7 & 173.5 & 197.2 C& 184.2 & 196.3 & 194.2 & 200.8 C& 235 & NM, $\overline{\tau}$ & \onlinecite{CePAS} \\
Sm  & rhombohedral    & 201.5 & 177.0 & 204.8 B& 187.8 & 202.3 & 198.5 & 206.4 B& 199 & NM & \onlinecite{SmPAS}\\
Gd  & hcp       & 202.9 & 178.0 & 203.3 C& 188.9 & 202.4 & 199.6 & 207.1 C & 230 & FM \\
W   & bcc       &  99.8 &  92.3 & 102.7 A&  96.6 & 101.9 & 100.6 & 103.4 A & 105 \\
Pt  & fcc       &  96.1 &  89.4 & 105.2 A&  93.4 & 101.1 &  97.4 & 101.3 A & 99+ \\
Pb  & fcc       & 188.9 & 167.2 & 202.3 A& 177.1 & 194.4 & 186.7 & 195.9 A & 200+ \\[1mm]
\multicolumn{11}{c}{Compounds} \\
MgO & rock salt & 117.8 & 108.7 & 145.4 C& 113.8 & 131.8 & 119.0 & 128.4 A & 130 & & \onlinecite{MgOPAS} \\
SiC & SiC-6H    & 136.1 & 123.4 & 145.2 A& 129.9 & 140.3 & 135.5 & 140.9 A & 140 & & \onlinecite{PASSiC6H}\\
Fe$_3$Al&D0$_3$ & 108.8 & 100.1 & 115.4 A& 104.9 & 113.2 & 109.2 & 113.6 A & 112+ & FM & \onlinecite{PASFe3Al} \\
ZnO & wurtzite  & 143.9 & 131.6 & 183.3 C& 138.2 & 162.3 & 144.6 & 156.9 B & 151+ & & \onlinecite{Brauer09} \\
CdSe& wurtzite  & 235.0 & 209.6 & 289.7 C& 222.2 & 258.9 & 235.3 & 254.2 C & 251,275 & & \onlinecite{weber} \\
PbSe& rock salt & 203.9 & 180.8 & 231.1 C& 191.6 & 215.2 & 202.4 & 214.7 B & 220  & & \onlinecite{polity}\\
\end{tabular}
\end{ruledtabular}
\label{tabL}
\end{table*}

\begin{table*}[htb]
\caption{Positron affinities (in eV) calculated according to various approaches (their designation is the same as in Table \ref{tabL}).
The last column gives experimental values when available.
If no reference is given, the value was taken from Refs. \onlinecite{Gidley88}, \onlinecite{Jibaly91} or \onlinecite{weiss} (see also Ref. \onlinecite{PAFelement}).
In columns GC and DG an evaluation of closeness of calculated and experimental values is given: A -- difference $\le0.2$ eV, B -- difference $\le0.5$ eV, C -- difference $>0.5$ eV.}
\begin{ruledtabular}
\begin{tabular}[t]{ l@{\extracolsep{20pt}} c c@{\extracolsep{1pt}} c@{\extracolsep{20pt}} c c c@{\extracolsep{1pt}} c@{\extracolsep{20pt}} c}
System & BN & GC &   & SG & DB & DG &   & experiment \\
\hline\\[-2mm]
\multicolumn{9}{c}{Elements} \\
Li  & $-7.65$ & $-7.42$ & & $-7.52$ & $-7.13$ & $-7.06$ & &  \\
C   & $-2.84$ & $-1.64$ &B& $-2.08$ & $-2.68$ & $-2.15$ &B& $-1.5$, $-2.0$\footnote{References \onlinecite{brandes,diamses}.}  \\
Ne  & $-18.94$ & $-16.11$ & & $-17.43$ & $-18.28$ & $-17.53$ &&  \\
Na  & $-7.68$ & $-7.34$ & & $-7.50$ & $-7.06$ & $-6.96$ & &  \\
Al  & $-4.55$ & $-4.38$ &B& $-4.46$ & $-4.22$ & $-4.17$ &A& $-4.1$ \\
Si  & $-7.08$ & $-6.50$ &B& $-6.81$ & $-6.65$ & $-6.52$ &B& $-6.2$\footnote{Reference \onlinecite{Si100pWF}.} \\
Fe  & $-4.31$ & $-3.74$ &B& $-4.03$ & $-4.12$ & $-3.98$ &C& $-3.3$ \\
Cu  & $-4.89$ & $-4.20$ &A& $-4.55$ & $-4.68$ & $-4.50$ &A& $-4.3$ \\
Zn  & $-5.34$ & $-4.78$ & & $-5.07$ & $-5.06$ & $-4.91$ & &  \\
Ge  & $-7.08$ & $-6.38$ & & $-6.75$ & $-6.63$ & $-6.47$ & &  \\
Nb  & $-3.99$ & $-3.62$ &A& $-3.81$ & $-3.75$ & $-3.65$ &A& $-3.8$ \\
Ag  & $-5.70$ & $-4.96$ &B& $-5.34$ & $-5.45$ & $-5.26$ &A& $-$5.2 \\
Sn-$\alpha$   & $-7.77$ & $-7.07$ & & $-7.44$ & $-7.24$ & $-7.08$ &&  \\
Sn-$\beta$    & $-6.46$ & $-5.99$ & & $-6.24$ & $-6.08$ & $-5.97$ &&  \\
Ce-$\alpha$   & $-4.76$ & $-4.36$ & & $-4.57$ & $-4.42$ & $-4.33$ &&  \\
Ce-$\gamma$   & $-5.94$ & $-5.58$ & & $-5.77$ & $-5.56$ & $-5.47$ &&  \\
Sm  & $-6.07$ & $-5.68$ & & $-5.88$ & $-5.68$ & $-5.58$ & &  \\
Gd  & $-6.11$ & $-5.75$ & & $-5.94$ & $-5.72$ & $-5.63$ & &  \\
W   & $-2.10$ & $-1.72$ &A& $-1.91$ & $-1.91$ & $-1.82$ &A& $-1.9$ \\
Pt  & $-3.93$ & $-3.31$ &B& $-3.64$ & $-3.77$ & $-3.61$ &A& $-3.8$ \\
Pb  & $-6.42$ & $-5.88$ &B& $-6.16$ & $-6.03$ & $-5.91$ &A& $-6.1$\footnote{Reference \onlinecite{PAFPb}.} \\[1mm]
\multicolumn{9}{c}{Compounds} \\
MgO & $-7.11$ & $-5.99$ &C& $-6.56$ & $-6.86$ & $-6.58$ &C& $-4.2$,$-2.4$\footnote{Reference \onlinecite{MgOPAF}.} \\
SiC-6H& $-5.51$ & $-4.88$ &B& $-5.22$ & $-5.22$ & $-5.07$ &C& $-4.4$\footnote{Reference \onlinecite{PAFSiC6H}.}  \\
Fe$_3$Al&$-4.10$&$-3.62$& & $-3.86$ & $-3.88$ & $-3.76$ & &  \\
ZnO & $-8.39$ & $-7.27$ & & $-7.85$ & $-8.05$ & $-7.79$ & &  \\
CdSe& $-9.21$ & $-8.28$ & & $-8.77$ & $-8.69$ & $-8.48$ & &  \\
PbSe& $-7.93$ & $-7.20$ & & $-7.59$ & $-7.50$ & $-7.34$ & &  \\
\end{tabular}
\end{ruledtabular}
\label{tabA}
\end{table*}

The positron lifetime is a fundamental characteristics which determines 
how long in average the positron lives in the material.
This quantity depends on whether the positron annihilates 
in the delocalized state (bulk annihilation) or in a localized one 
(annihilation in a defect).
In the former case, the positron lifetime represents a bulk property, 
whereas the lifetime corresponding to a defect is case dependent.
The positron lifetime (for defects and bulk) can be both 
measured and calculated by using Eq.~(\ref{eq:lambda}).
Before discussing calculated positron bulk lifetimes 
and their correspondence to measured values, 
it is useful to mention some aspects of positron lifetime experiments.
Measuring positron lifetimes for real samples is a well established procedure.\cite{krause}
However, there is a rather big scatter between measured positron lifetimes 
reported in literature (see e.g. Ref. \onlinecite{perexp03}).
Differences of the order of 10 ps for bulk lifetimes are not rare, 
even when only recent and reliable measurements are considered.\cite{campillo}
This problem complicates the comparison of theory with experiment 
and rises questions regarding the precision of lifetime measurements though the statistical precision of bulk lifetime measurements is typically around 1 ps only.
Standard positron lifetime experiments use the so-called sandwich setup 
in which the positron source (usually based on the $^{22}$Na isotope) 
is ``wrapped'' in a thin foil and is surrounded by two identical pieces of the
samples to be examined.
The source is the origin of additional lifetime components, 
which appear in the lifetime spectrum of the 
measured samples.\cite{sourceEldrup,sourceStaab,sourceDundee}
The elimination/subtraction of such source components
(even if their intensities are not large)
is generally a non-trivial task and various approaches for
this procedure can be taken.\cite{sourceEldrup,sourceStaab,sourceDundee,melt}
In general, source lifetimes' admixtures are the origin of uncertainty in lifetime measurements.
For example, McGuire and Keeble\cite{sourceDundee} have shown that using three different wrap foil materials for the source
results in differences up to 6 ps in bulk lifetimes of Al, Ni, Zr, and Pb and up to 9 ps for vacancy lifetimes in these
same materials.
Moreover, in materials where the source and the sample lifetimes are close, further complications affect
the data analysis and makes it prone to systematic errors.
Lifetime measurements with positron beams may circumvent the source correction problem,
however, the lifetime resolution function of current positron beam spectrometers 
is usually worse compared to standard setups.
Finally another issue for measurements of bulk lifetimes is that samples 
may contain a small amount of defects -- even if the samples are ``well annealed'' -- 
which cannot be detected as separate lifetime components in experiments.
The amount of such defects (dislocation lines, stacking faults, impurity related defects, etc.) is not normally checked with other methods.
Therefore, a broader effort in the positron experimental community -- 
such as the collective lifetime study\cite{lifetimejapan} by 
12 Japanese positron laboratories 
-- might be necessary to tackle these problems.

There are several publications\cite{seeger,krause,perexp03,sourceEldrup} which review and summarize experimental bulk and sometimes also monovacancy lifetimes.
We shall use these review articles as a background for the comparison of our calculated positron lifetimes 
with the measured ones. We will also discuss some specific cases in more detail.
Our choice of experimental lifetimes for elements to be compared 
with the present calculations follows to some extent 
the compilation presented by Campilo Robles {\em et al.}\cite{campillo}
Table \ref{tabL} contains in the last column the references where the corresponding lifetimes were taken from. 
If no reference is explicitly specified, we took values from Refs. \onlinecite{seeger,perexp03,sourceEldrup}. 
Except for the alkali metals and solid neon, we have 
considered only data after 1975 since the evaluation of experimetal data
have become more reliable after this period.\cite{campillo} 
In Table \ref{tabL}, we give some measured lifetime values in the form ``X+'', 
where X is the minumum bulk lifetime detected,
since the bulk lifetime can appear slightly longer because of the possibility of incomplete 
source corrections and the fact that the bulk lifetime to be measured is usually shorter than the source lifetime components.\cite{Eldrup}

Table \ref{tabL} shows that the AP theory underestimates systematically the positron lifetime in real materials. The BN, SL and DB enhancement factors reduce the overestimation of annihilation rates obtained with AP.  Our AP and BN lifetimes are in excellent agreement with the corresponding LDA values calculated by Takenaka and Singh within the all-electron linearized augmented plane wave method.\cite{Takenaka}
The BN, SL and DB enhancement factors reduce the overestimation of annihilation
rates obtained with AP.
Therefore, the effect of the gradient correction turns to be smaller for SG\cite{ggaboro} and for DG.
In the case of the DG approach, the overall agreement between experiment and the GGA is very good when $\alpha=0.05$, particularly when considering that the calculated values
result from a self-consistent, full potential and all-electron approach.
One of the best result is given by Al which was problematic in the old GGA scheme.\cite{gga1}
Transition metals and semiconductors also give excellent results within the new GGA scheme.
However, some disagreement between the QMC (DG) and experiment is exhibited by the alkali metals. Nevertheless, lifetime experiments in the alkali metals are rather old and new measurements are needed in order to confirm these discrepancies. Another problem is the long positron thermalization time \cite{thermalization} which complicates the lifetime analysis in the alkali metals.
In order to evaluate the deviation of calculations and experiment we give in columns GC and DG a measure expressed by characters A, B and C. 
This measure is based on the root mean square deviations determined using available experimental data where we considered rather newer than older data removing also unrealistic results. 
A, B and C mean, respectively, a deviation smaller than 5 ps, a deviation between 5 and 10 ps and a deviation larger than 10 ps.
The DG approach compares somewhat better to experiment than the original gradient scheme though one should be aware of experimental uncertainties, as discussed above.

Table \ref{tabA} presents the calculated positron affinities.
The DB $V^{ep}$ correlation potentials reduce the overestimation of the electron-positron
correlation obtained with AP and give very good agreement between experiment and theory, particularly for Al, Cu, Nb, Ag, W, Pt and Pb at the LDA level. 
For the DG case the GGA positron correction in these metals becomes comparable to the experimental error bars of the order of $0.1$ eV since $\alpha$ has been reduced to $0.05$.
The level of correspondence between experimental and calculated values (GC and DG results) is also evaluated and specified in Table \ref{tabA} by A, B and C letters.
One can see that the DG approach provides slightly better agreement with experiment compared to GC. 
It should be also taken into account that the uncertainty of some experimental values can be quite large.

We shall now comment on some materials in more detail.
In Table \ref{tabL} we also report the positron lifetime in solid Ne. This material can be used as a moderator for producing slow positron beams.\cite{gullikson1986}
The efficiency of such moderator is superior to the standard moderator based on tungsten.
The positron lifetime measurement in solid neon performed by Liu and Roberts\cite{liu1963} in 1963 was performed with a rather poor resolution. Nevertheless, it shows some discrepancy  with the present GGA theory which could support the hot-positron
model.\cite{gullikson1986,mills1986}
In fact, our theory considers a positron wave function in the lowest energy state while in the hot-positron model, most positrons do not thermalize and are in excited states. We should also mention that some DFT failures in correctly describing the positron correlation potential can contribute to the lifetime discrepancy as well.\cite{puska1992}
The positron affinity of neon shown in Table \ref{tabA} is a large negative number compared to the positron affinities for metals and semiconductors as in Ref. \onlinecite{puska1992} but the GGA  corrects some exaggerated correlations effects present in the LDA.

Elemental semiconductors (C, Si and Ge) exhibit a good agreement of measured and calculated lifetimes for the new GGA scheme, especially for C and Si.
Concerning positron affinities, positron reemission from the (100) surface of diamond has been extensively studied by Brandes and Mills.\cite{brandes}
The experimental affinity values given in Table \ref{tabA} were determined on the basis of this work considering also electron work functions of clean and monohydride (100) diamond surfaces as determined by Diederich {\em et al.}\cite{diamses} taking into account corresponding (100) band bendings of IIa type (N-doped) diamond examined in Ref. \onlinecite{brandes}.
The agreement of gradient corrected values with experimental ones is encouraging.

The calculated positron affinity (GC, DB and DG) values of silicon agree relatively well with the recent experimental study of the Si(100) surface by Cassidy {\em et al.},\cite{Si100pWF} where its positron work function $\phi^+$ = 0.8 eV was determined using positronium formation potential measurement.
It is worth noting that $\phi^+$ (for (100) surface) is positive in contrast to most of other materials, which is probably related to rather loose atomic arrangement of Si, resulting in quite a low positron level with respect to vacuum. 
Earlier measurements carried out on Si provided negative\cite{SipWorkF} and nearly zero\cite{schultz} values of $\phi^+$ for (100) and (111) surfaces, respectively. 
In any case, determining the positron affinity for semiconductors is a difficult task that can be affected by various surface effects including reconstruction.
An accurate measurement of the electron work function is also necessary, which is demonstrated by somewhat different results of early\cite{SieWorkF} ($4.9$ eV) and recent\cite{SieWFnew} ($5.4$ eV) experiments.
A positron beam study\cite{MOS} on a metal-oxide-system with a silicon substrate indicated a value of about 5 eV.

Positron lifetime allows to study an important phase transition in tin
at $T_c$ = $13.2 \,^{\circ}{\rm C}$ (Ref. \onlinecite{SnLifetime}).
The two different phases of tin are the white metallic $\beta$-tin
with tetragonal (A5) structure and the grey, semiconducting
$\alpha$-tin with diamond structure.
The $\alpha$-tin has recently attracted particular interest because
in the presence of uniaxial strain, it can become a strong topological
insulator.\cite{topins}
The transition $\beta$ to $\alpha$ is also accompanied by
a large increase in volume of about $27\,\%$
which results in an increase of lifetime of about $90$ ps.
As shown in Table \ref{tabL}, the present GGA scheme gives the best
agreement with the experiment.

Table \ref{tabL} shows that the calculated positron lifetime could be also sensitive
to the $\alpha-\gamma$-transition in fcc Ce.
However, the experimental lifetime does not change much
and it is closer to the new GGA scheme lifetime for $\gamma$-Ce.
This is probably an indication that positrons always annihilate
in patches of $\gamma$-Ce embedded in $\alpha$-Ce though only the mean lifetimes are available for both phases.\cite{CePAS}
This hypothesis is in fact validated by Table \ref{tabA}, which shows
that patches of $\gamma$-Ce embedded in $\alpha$-Ce
produce potential wells with a depth larger than 1 eV.
The new GGA scheme gives the best overall agreement
with the rare earth metals. However, the accuracy of the theory
could be improved by considering
temperature dependent DFT calculations in which vibrational,
electronic and magnetic free energies are
taken into account.\cite{Jarlborg1997}
When partially filled $f$-orbitals are involved, the ground state predicted by the DFT
clearly places the $f$-electrons in narrow bands piled at the Fermi energy $E_F$,
interacting only weakly with other electrons. In sharp contrast, however, signatures
of $f$-bands are often found in spectroscopic measurements not at $E_F$, 
as band theory predicts, but several eV's above or below the $E_F$ depending on the nature of the spectroscopy.\cite{Jarlborg2012}
Thus it is interesting to examine the extent to which $f$-electrons 
in the ground state can contribute to Fermi-surface-related properties.
Interestingly, if $f$-electrons at or near the Fermi level positron 
annihilation positron-annihilation can detect them.\cite{vasumathi1999}

Concerning oxide materials, we have focused on MgO and ZnO, which are involved 
in several promising nanotechnology applications.\cite{pacchioni,ozgur}
MgO has been examined many times by positrons (see e.g. Ref. \onlinecite{MgOPAS} and references therein).
Early investigations have been challenged by the following puzzle: 
Measured MgO lifetimes were over 150 ps whereas calculated bulk lifetimes 
yield much shorter values. 
This problem has been explained by showing that unintentional 
MgO doping\cite{MgOdoping} (e.g. by Ga) causes creation of Mg vacancies, 
which increases the lifetime.
Currently, the accepted bulk positron lifetime for MgO is 130 ps and 
it agrees very well with the SG and DG values presented in Table \ref{tabL}.
The positron affinity of MgO has been estimated from the positronium 
formation potential by van Huis {\em et al.}\cite{MgOPAF}
As shown by Table \ref{tabA}, the reported values are too small in magnitude 
compared to calculated ones.
ZnO is perhaps even more interesting since its bulk positron lifetime is still 
debated in the positron community.
The situation might be somewhat similar to that for MgO.
Unintentional doping with H (Ref. \onlinecite{Brauer09}) 
and Li (Ref. \onlinecite{Johansen11}) 
could play an important role.
The present calculations following 
the SG and DG approaches suggest that the ZnO bulk lifetime is close to 160 ps.

Among technological relevant composite materials, silicon carbide is a wide band gap semiconductor used for high temperature applications,\cite{casady} which exhibits polytypism.
Probably the most frequently studied polytype is the 6H one for which we have performed positron calculations.
The positron lifetime is sensitive to the gradient correction in a smaller extent compared to oxides discussed above.
All gradient corrected lifetimes agree reasonably well with the experimental 
value $\sim$140 ps (Ref. \onlinecite{PASSiC6H}).
The experimental positron affinity of SiC-6H reported in Ref. \onlinecite{PAFSiC6H} 
also matches satisfactorily calculated counterparts, the gradient corrected values being closer.
It is also instructing to discuss Si and C (diamond) results in relation to SiC.
The Si (C) lifetime is apparently longer (shorter) than the SiC value due to a looser (more compact) atomic arrangement of Si (C) atoms though atomic arrangement type is similar for Si, C and SiC.
The positron affinity for C, SiC and Si follows the lifetime trend: the magnitude of $A^+$ increases in this series as the atomic arrangement becomes looser, which, roughly speaking, corresponds to the positron level getting deeper through the series.

In the case of II-VI compound semiconductors, the atomic spherical approximation (ASA)
used in the LMTO method\cite{bdj} influences the calculated positron lifetime. For instance, 
the positron lifetime for CdSe calculated within the ASA (Ref. \onlinecite{weber}) is few ps shorter than 
the corresponding result without shape approximations given in Table \ref{tabL}.
The lifetime of positrons implanted into bulk CdSe with 2 keV has been measured and an experimental 
value of 275 ps has been found while the corresponding lifetime measured in the 6-nm CdSe sphere 
was 251 ps (Ref. \onlinecite{weber}). The latter value is in better agreement with the new 
GGA schemes reported in Table \ref{tabL}.
The positron affinity can be studied as well and the ASA error is of the order 1 eV.
Using the experimental electron work function of 6.62 eV (Ref. \onlinecite{nethercot}) and the theoretical 
positron affinity of 8.12 eV based on our calculations, the positron work function can be deduced to the order 
of 1.5 eV indicating that positron can be trapped inside the nanocrystal. However, the positron potential well 
may deepen near the surface of the sample and the positron can therefore form surface states.\cite{Eijt}

CdSe  has a  wurtzite structure while PbSe adopts a rock salt crystal structure and has a more ionic nature.
As a result the positron affinity is smaller in amplitude in PbSe compared to CdSe as shown in Table \ref{tabA}.
Our calculations based on the new GGA scheme reveal a positron lifetime for bulk PbSe of 215 ps, 
which is in good agreement with the experimental value 220 ps (Ref. \onlinecite{polity}). 
Moreover, positron lifetimes ranging from 340 and 380 ps observed at PbSe nano crystals demonstrate 
the existence of positron surface states.\cite{apl2013} Therefore, positron annihilation can 
be used as an advanced characterization tool to unravel many novel properties 
associated with the surface physics and chemistry of nanocrystals.

In order to complement the calculations for composite semiconductors, we have also computed positron characteristics for intermetallic compound Fe$_3$Al.
To the best of our knowledge, the positron affinity for this material has not been measured.
The experimental positron lifetime ($\sim$112 ps; Ref. \onlinecite{PASFe3Al}) 
agrees well with all calculated gradient corrected values.

We have also examined the influence of the choice of the XC potential for electrons.
The XC functional used in the electronic structure calculations to produce Tables \ref{tabL} 
and \ref{tabA} was within the LDA as in Ref. \onlinecite{gga1} to facilitate comparisons with existing literature.
If the LDA XC potential is replaced by the GGA potential\cite{ePBEXC} the positron lifetime in Table \ref{tabL} 
does not change significantly (usually less than 1 ps).
However, the situation is different for the positron affinity, where the gradient correction on the 
XC potential can produce shifts as large as 0.5 eV.
Nevertheless, usual applications of the positron affinity consist in finding 
the affinity difference between two phases (e.g. a matrix and an embedded cluster\cite{Nagai}), 
thus the affinity shifts due to the electron GGA mostly cancel.
Interestingly, refined experimental affinities values could be also very useful 
to better understand adequacy and known deficiencies of the electron LDA and 
GGA XC functionals.\cite{bmj90}

\subsection{Details of positron gradient correction}

\begin{figure}[tbh]
\includegraphics[width=8.5cm, clip]{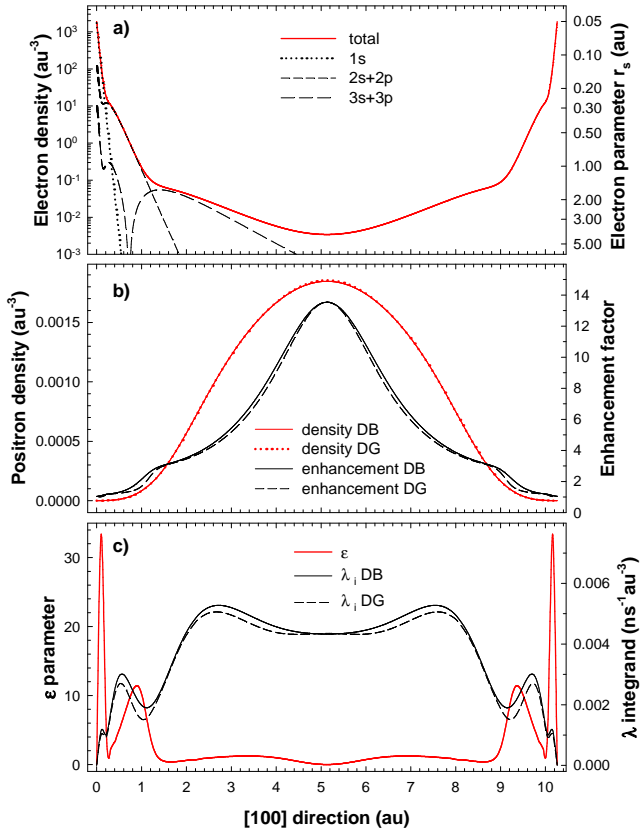}
\caption{(Color online) One dimensional profiles of (a) the electron density, (b) the positron density and enhancement factor, and (c) the $\epsilon$ parameter and $\lambda$ integrand along [100] direction in Si for DB and DG approaches.}
\label{figSi1D}
\end{figure}

The effect of positron GGA at the microscopic level will be now illustrated for Si and Na,
which are examples of a semiconductor and a metal having open and close-packed structures, respectively.
Fig. \ref{figSi1D} shows one-dimensional (1D) profiles of several electron and positron quantities of interest together with the influence of the gradient corrections. The DB and DG approaches were employed for LDA and GGA positron calculations, respectively. In Fig. \ref{figSi1D}a, the total, WIEN2k self-consistent electron density is plotted along [100] direction. The plot shows the density on a line connecting two Si atoms located at the positions 0 and 10.26 a.u.
This line goes through the center of large interstitial space located between atoms at line ends.
Purely atomic orbital densities are added to the plot to illustrate where these orbitals yield the dominant contribution to the total density. In particular, the 1$s$ orbital dominates close to nucleus while farther from the atom, 2$s$ and 2$p$ orbitals become important and finally, beyond a distance of about 1.6 a.u. from the nucleus, the 3$s$ and 3$p$ orbitals prevail but lose their atomic character because of strong charge transfer and hybridization effects. These atomic shells are rather well delineated by
the $\epsilon$ parameter as shown in Fig. \ref{figSi1D}c. This parameter has clear maxima in 1$s$ and
2$s$+2$p$ regions. There are also smaller maxima in the 3$s$+3$p$ region, but these features are more related to the interstitial charge than to particular atomic orbitals.
The enhancement factors for the DB and DG approaches are presented in Fig. \ref{figSi1D}b. The DB enhancement exhibits a reduction in the regions of $\epsilon$ maxima (except in part in the 1$s$ region where the enhancement is close to 1 since the $r_s$ parameter is very small).
Consequently, the DG $\lambda$ integrand (shown in Fig. \ref{figSi1D}c) is reduced compared to the DB one.
This effect results in a smaller positron annihilation rate for the DG case and thus a longer lifetime as expected.
The positron density (Fig. \ref{figSi1D}b) remains almost unaffected by the gradient correction. Thus, the present Si example demonstrates the importance of the gradient correction in the interstitial space of open structures, in addition to core electron regions.
Some noticeable jumps of $\epsilon$ shown in Fig. \ref{figSi1D} may lead to sharp features in the electron-positron correlation potential $V^{ep}$, which are absent in a non-local density approach called weighted density approximation\cite{WDA} (WDA). 
Stachowiak and Boro\'nski have noticed that the WDA better describes some of such inhomogeneities.\cite{stachowiak}

\begin{figure}[tbh]
\includegraphics[width=8.5cm, clip]{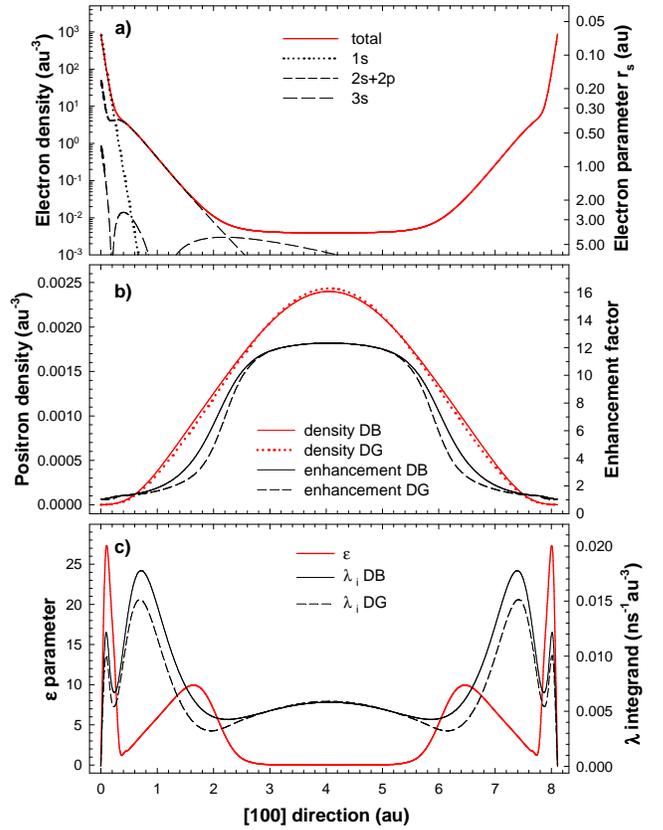}
\caption{(Color online) One dimensional profiles of (a) the electron density, (b) the positron density and enhancement factor, and (c) the $\epsilon$ parameter and $\lambda$ integrand along [100] direction in Na for DB and DG approaches.}
\label{figNa1D}
\end{figure}

The positron gradient correction for the core electrons is further illustrated with the example of Na shown in Fig. \ref{figNa1D}. The plot of the self-consistent electron density and of the atomic orbitals (Fig. \ref{figNa1D}a) closely resembles the corresponding Si plot in Fig. \ref{figSi1D}a.
We note that the line along which the density is plotted connects two atoms along the [100] direction.
The differences between Si and Na are that Na has no 3$p$ electrons and that the total electron density is rather flat in the interstitial region dominated by the 3$s$ delocalized, conduction electrons.
Both Si and Na have similar minimum electron densities in the interstitial space.
The maxima of the parameter $\epsilon$ of Na correspond (as in the case of Si) to 1$s$ and 2$s$+2$p$ electrons (see Fig. \ref{figNa1D}c). There are, however, no $\epsilon$ maxima in the interstitial space because of the almost constant electron density in this region.
Thus, one can expect that there will be no enhancement reduction in the interstitial space.
Indeed, the enhancement factor plot in Fig. \ref{figNa1D}b confirms this behavior.
On the other hand, there is a strong reduction of the enhancement factor in the 2$s$+2$p$ region, which is also reflected by the $\lambda_i$ quantity in Fig. \ref{figNa1D}c.
This trend yields a large reduction of the annihilation rate and consequently leads to a prolongation of the positron lifetime (by 21 ps) due to the gradient corrections.
Sodium thus represents an example of material where the positron gradient correction is important only in the region of core electrons.

\begin{figure*}[htb]
  \includegraphics[width=6.5cm]{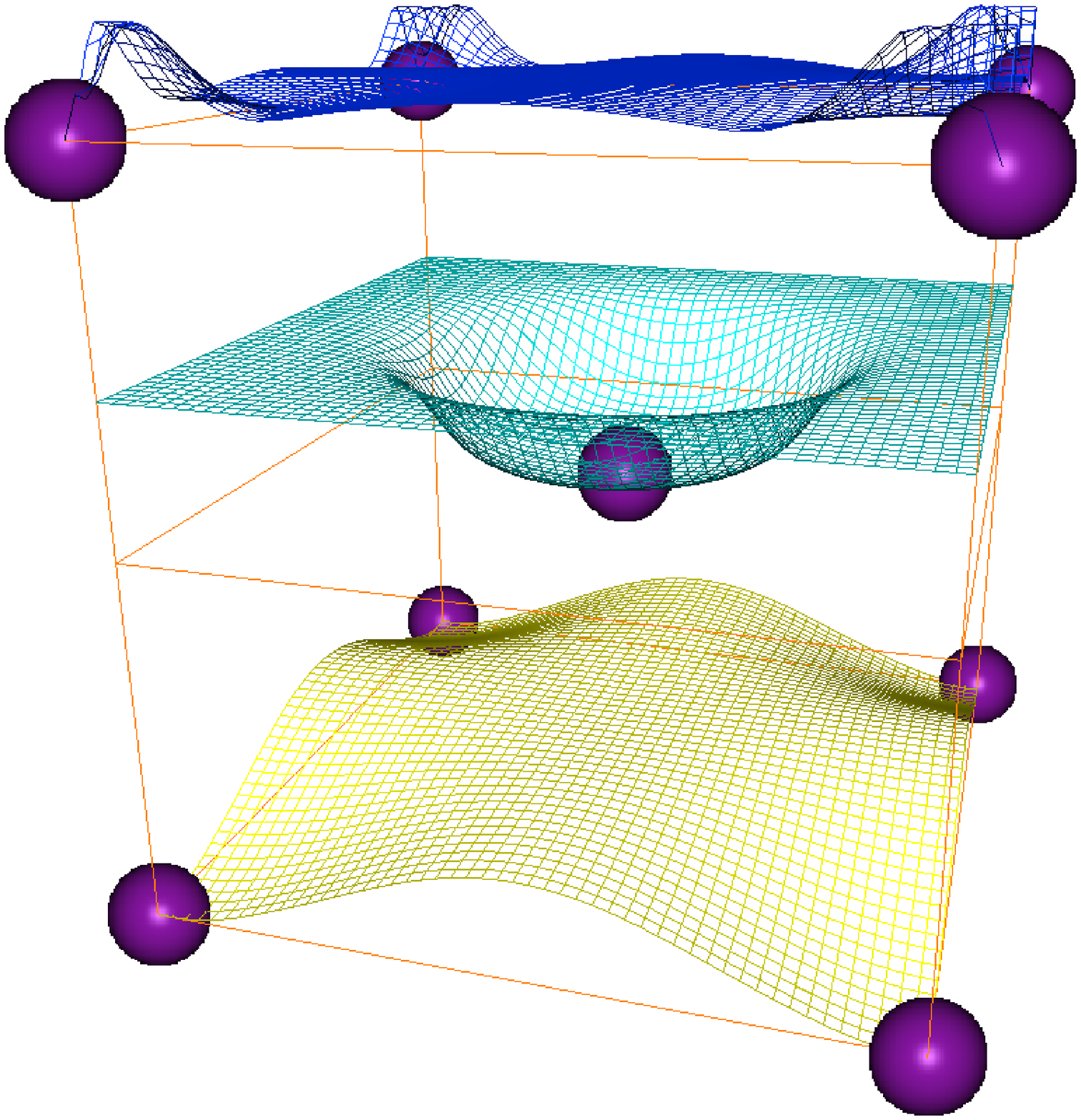}%
  \hspace{10mm}\includegraphics[width=6.0cm]{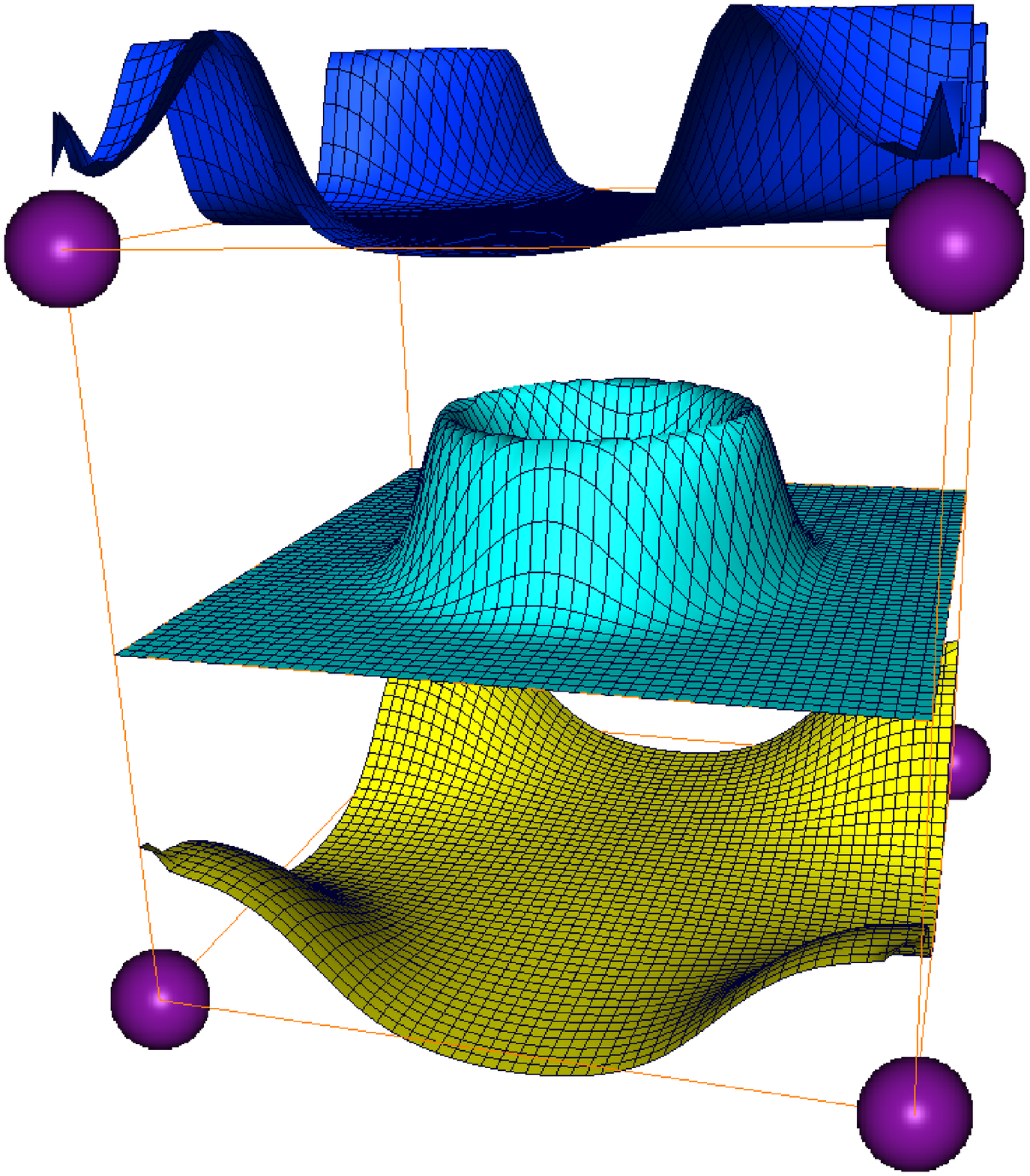}
  \caption{(Color online) Left: plots of the positron density (lower plane, yellow (light) color), enhancement factor (middle plane, cyan (medium) color) and annihilation rate integrand (upper plane, blue (dark) color) obtained using the DG approach for Na (bcc structure). Atoms are represented by violet (dark) spheres.
The right picture shows the corresponding relative changes due to the gradient correction (e.g. for the positron density the change is defined as
$(\rho_{DB}^+ - \rho_{DG}^+)/\rho_{DB}^+$).
The ranges of relative changes are as follows: $\rho^+$ ($-1.5$\%, $+12.1$\%),		
$\gamma$ ($0.0$\%, $+42.8)$\% and	$\lambda_i$  ($-1.5$\%, $+34.1$\%).
}
\label{figNa}
\end{figure*}

\begin{figure*}[htb]
  \includegraphics[width=6.5cm]{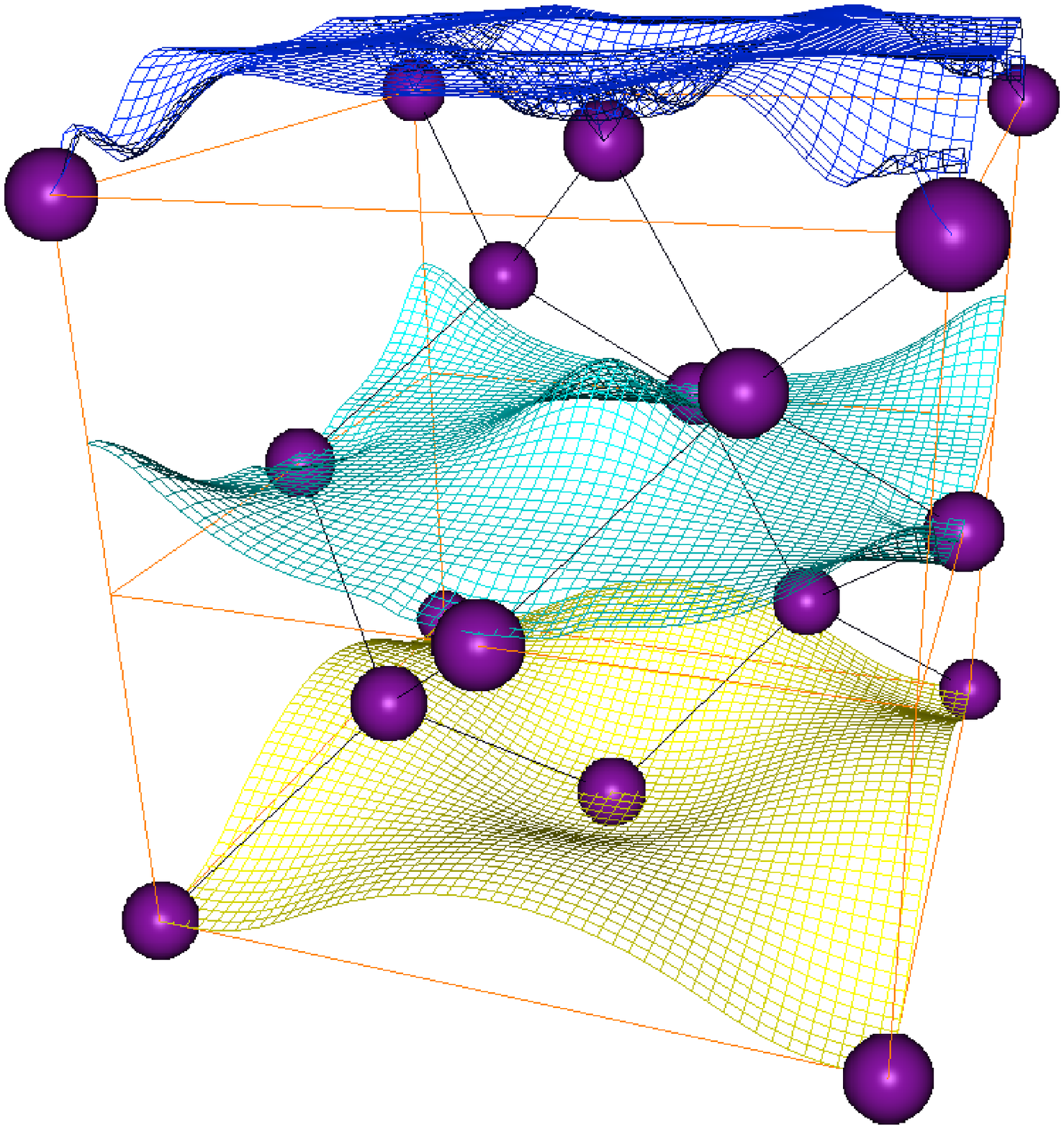}%
  \hspace{10mm}\includegraphics[width=6.3cm]{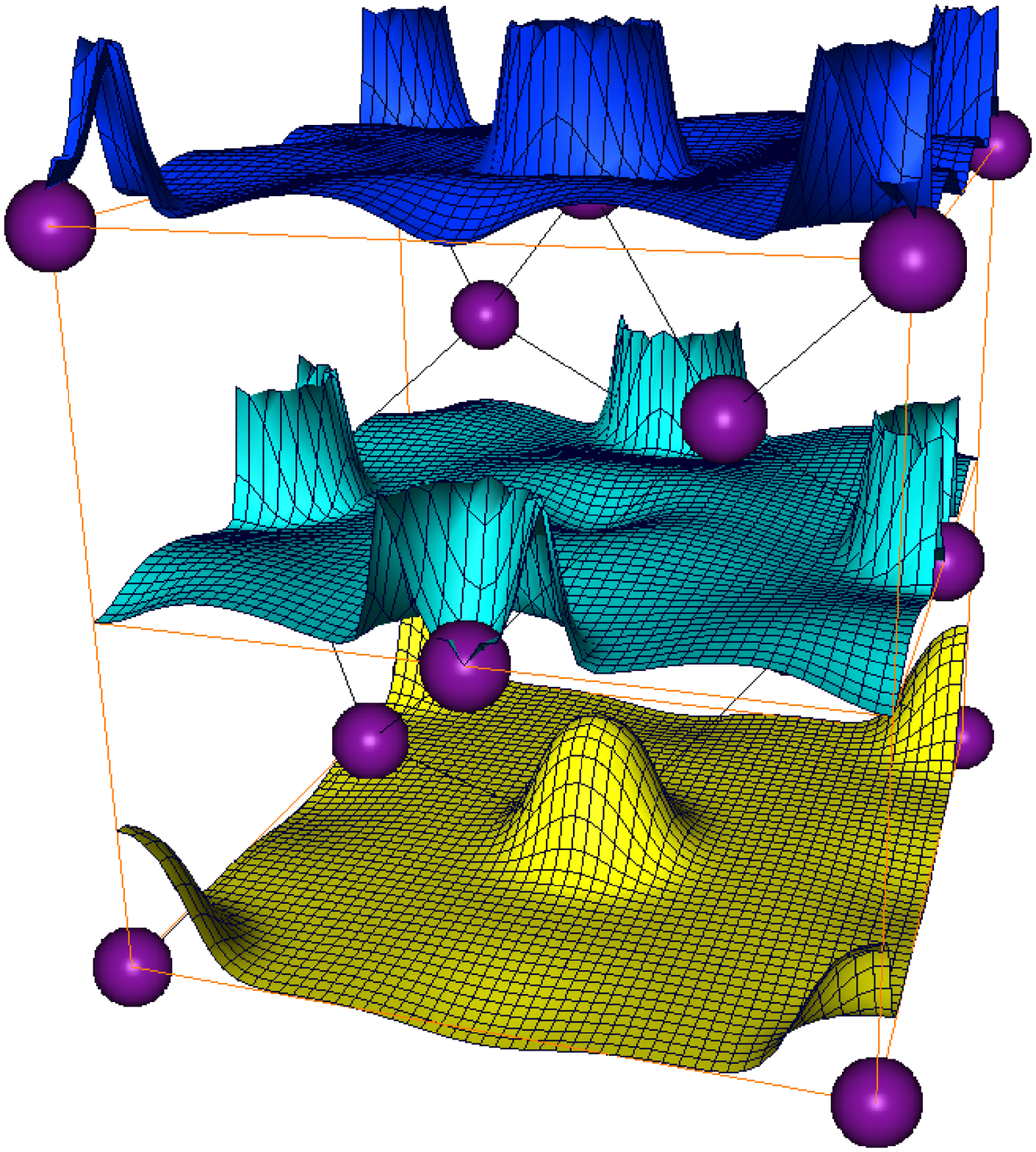}
  \caption{(Color online) Plots of $\rho^+$, $\gamma$ and $\lambda_i$ and their relative changes for Si (diamond structure).
The organization of the figure is the same as for Fig. \ref{figNa}.
The ranges of relative changes are as follows: $\rho^+$ ($-0.5$\%, $+5.2$\%),
$\gamma$ ($0.0$\%, $+29.7)$\% and	$\lambda_i$  ($-0.5$\%, $+24.9$\%).}
\label{figSi}
\end{figure*}

The above considerations are helpful to understand the positron gradient correction, but are based on one dimensional profiles cutting through the crystal of examined materials.
Therefore, we further discuss by showing the quantities of interest on two-dimensional crystal planes.
Figures \ref{figNa} and \ref{figSi} show the plot of the positron density, enhancement factor and the integrand of the annihilation rate ($\lambda_i$) in several planes cutting the bcc Na and diamond Si structures, respectively.
The relative changes due to the gradient correction are also shown in the right panel of the figures.
In both figures, the positron density exhibits nearly zero values close to the nuclei and reaches its maximum in the interstitial region. This is the expected behavior since positrons are strongly repelled from positively charged nuclei. Besides, the enhancement factor is slightly above 1 near nuclei (see Eq. (\ref{eq:gamma})) since the electron density is very large here (and $r_s$ is thereby small).
In the interstitial region, where the electron density is low,
$\gamma$ is reaching its maximum ($>10$).
These trends correlate well with those illustrated in Figs. \ref{figSi1D}b and \ref{figNa1D}b.
The 2D plots of $\lambda_i$ identify which parts of the crystal contribute to the annihilation rate.
For sodium, the largest local contribution $\lambda_i$ is from regions near nuclei
while in the case of silicon, the largest contribution to the annihilation rate originates from the interstitial space (this trend is also consistent with the picture that positrons reside and annihilate primarily in the interstitial regions).

The Na case illustrated in Fig. \ref{figNa} exhibits large relative changes due to the gradient correction around atoms (in the core region), but there is almost no change in the interstitial space far from atoms, as one can see from the enhancement plot.
Thus, the lifetime change caused by the gradient correction is mainly due to the core electron enhancement change -- as we already discussed above -- and also confirmed by the $\lambda_i$ plot.
The relative change of the positron density is rather small compared to the changes of $\gamma$ and $\lambda_i$ and the largest modification occurs in the core region.
In contrast to Na behavior, the enhancement changes in Si, shown in Fig. \ref{figSi}, are important in the whole unit cell though some large effects still occur in the core region.
The behavior of $\lambda_i$ confirms that the effect of the gradient correction is important in the whole cell.
Moreover like for Na, the relative changes of the positron density are rather small and the main changes are in the core region.
These conclusions for Si thus also confirm those made above when discussing
the one dimensional profiles shown in Fig. \ref{figSi1D}.

\subsection{Vacancy calculations}

In addition to bulk positron characteristics computations, we have calculated positron annihilation rates and binding energies for monovacancies in  Al, Si and Cu.
At this stage, we do not consider lattice relaxations due to the defect itself and due to positron induced forces, but it is well known that these two effects compensate to quite a large extent.\cite{Makkonen06}
More precisely, positrons localizing at vacancies can influence the electron density.\cite{tdft,ATSUP2,gilgien,torrent} 
Nevertheless, since we consider here vacancies which trap the positron only moderately, we will neglect the modifications of the electronic structure in the defect due to the positron.
The results of positron lifetime calculations are presented in Table \ref{tabV} for all approaches to electron-positron correlations examined above.
The table also contains positron binding energies to defects.

\begin{table*}[tbh]
\caption{Positron lifetimes ($\tau$) and positron binding energies ($E_b$) calculated according to various approaches to electron-positron correlations for a single vacancy in Al, Si and Cu.
$E_b$'s for the BN approach are identical with the corresponding AP and SL ones (the same $V^{ep}$ is used).
The last column gives experimental lifetime values extracted from the collection of lifetime data in Ref. \onlinecite{perexp03}.
}
\begin{ruledtabular}
\begin{tabular}[t]{@{\ }l cc c cc c cc cc cc c}
Element & \multicolumn{2}{c}{BN}& AP& \multicolumn{2}{c}{GC}& SL&
\multicolumn{2}{c}{SG}& \multicolumn{2}{c}{DB}& \multicolumn{2}{c}{DG}& Experiment \\
    & $\tau$& $E_b$& $\tau$& $\tau$& $E_b$& $\tau$& $\tau$& $E_b$& $\tau$& $E_b$&
      $\tau$& $E_b$& $\tau$ \\
    & (ps)&(eV)& (ps)& (ps)&(eV)& (ps)& (ps)&(eV)& (ps)&(eV)& (ps)&(eV)& (ps) \\
\hline\\[-2mm]
Al& 234.1&1.03& 204.1& 222.3&0.91& 217.2 & 227.0&0.98& 228.3&0.91& 233.6&0.89& 237--244 \\
Si& 241.8&0.36& 211.2& 253.5&0.42& 224.8 & 245.0&0.39& 238.5&0.32& 249.1&0.34& 270--273 \\
Cu& 162.7&0.94& 146.7& 183.9&1.01& 154.6 & 173.5&0.98& 158.7&0.79& 168.7&0.81& 180$\pm$5 \\
\end{tabular}
\end{ruledtabular}
\label{tabV}
\end{table*}

When our self-consistent results are compared to with those of Campillo-Robles {\em et al.},\cite{robles2013} the main effect of the full potential calculation is to decrease the value of the positron lifetime obtained within the ASA.
The reason for the differences between the full potential and the ASA can be understood  as follows.
Within the ASA, a vacancy is approximated by an empty sphere.
The localization of the positron wave function in the empty sphere is stronger than in the actual interstitial region.
Therefore, this stronger positron localization increases the positron lifetime.
All our LDA values are systematically lower than the corresponding experimental values,
but the agreement with the experiment is improved  with GGA and could be further improved by considering appropriate lattice relaxations mentioned above.

Experimental lifetime values given in Table \ref{tabV} are mostly based on the review in Ref. \onlinecite{perexp03}. 
We selected experimental results with the source correction subtracted.
As in the case of the bulk positron lifetime, vacancy lifetimes are scattered to some extent (except for Cu) and depend on experimental setup and data evaluation procedure (including the source correction).
In the case of the Si vacancy, the situation is complicated by the possibility of its various charge states which might exist in measured samples.\cite{Makinen89,Dannefaer86}
In our calculations, we considered the neutral charge state only.
Furthermore, vacancy-impurity complexes introduced 
by either unintentional or intentional doping of Si 
(e.g. Czochralski-grown samples contain oxygen atoms) 
can modify the lifetime spectrum, as discussed e.g. in Ref. \onlinecite{Makinen89}.
Regarding the studied fcc metals Al and Cu, the situation can be complicated 
by the existence of dislocations and stacking faults.
These lattices imperfections can bind single vacancies and therefore can 
affect the measured lifetimes as well.

The positron binding energy to defects is a very important quantity, but it is hardly accessible experimentally.
So far only defects which exhibit positron detrapping were investigated experimentally in order to evaluate their positron binding energy (see e.g. Ref. \onlinecite{Arutyunov13}).
On the basis of Table \ref{tabV} we can see that a positron traps quite weakly in the Si vacancy in contrast to metallic vacancies both in Al and Cu where the binding energies are about 2.5 times higher.
Calculations of binding energies for various defects allows to determine whether positrons may ever get trapped in such defects.
For instance, oxygen vacancies in oxides are often found not to trap positrons.\cite{Brauer09}

\section{Conclusion}  \label{End}

We have calculated positron characteristics in selected, representative solids based on reliable DFT electronic structure calculations taking into account highly precise electronic charge transfers due to an electron self-consistent full potential.
The positron energetics has been monitored by calculating the positron affinities.
The new LDA scheme based on accurate QMC simulations\cite{qmc} 
improves systematically the affinities obtained with the other LDA schemes.
However, the LDA does not take into consideration charge inhomogeneities due to non metallic charge distribution and the effect of the electron-nuclear interaction which is disrupting the pile-up of electronic charge around the position of the positron.
Therefore, we show that gradient corrections to the LDA are still needed in such circumstances despite their intensity is reduced in comparison to the original GGA scheme.\cite{gga1}
Our study has used well converged electronic structures without any shape approximations for the charges and the potentials, controlling also important numerical parameters of the WIEN2k calculations performed, and has confirmed preliminary results of Boro\'nski\cite{ggaboro} suggesting that the GGA-PHNC was a right step towards an improved GGA scheme for positron states in materials.
Especially the positron lifetime is a very sensitive measure of any GGA scheme for positrons as it is determined directly from the electron density.
At the moment, it is difficult to decide -- by making comparisons with available experimental positron lifetime and affinity data -- which is the best GGA approach among the GGA-PHNC (SG) and the GGA-QMC (DG). Thus, more precise experiments are needed to sort out this important matter.
The present GGA scheme could be further improved by extracting the parameter $\alpha$ from many body physics. 
This more accurate determination of $\alpha$ may also reveal a gentle dependence of the local density, thus $\alpha$ could become function of the local $r_s$. 
A new WDA scheme based on QMC data could be also another route to capture non-local effects of the density.

\begin{acknowledgments}
The authors would like to thank M.~Eldrup, D.~Keeble and I.~Proch\'azka for helpful discussions concerning various aspects of positron lifetime measurements. 
Discussions with G. Brauer, S. W. H. Eijt, and A. P. Mills, Jr. about measurements of positron work functions and affinities are greatly appreciated.
N.D.~Drummond is acknowledged for providing us with numerical results of QMC calculations.
This research used resources of the National Supercomputing Center IT4Innovations (Czech Republic), which is supported by the Op VaVpI project number CZ.1.05/1.1.00/02.0070.
B.B. is supported by the US Department of Energy, Office of Science, Basic Energy Sciences Contract No. DE-FG02-07ER46352. He has also benefited from Northeastern University's Advanced Scientific Computation Center (ASCC), theory support at the Advanced Light Source, Berkeley, and the allocation of computer time at NERSC through Grant No. DE-AC02-05CH11231.
\end{acknowledgments}

\appendix*
\section{Computational details} 

\subsection{Positron calculations}  \label{APC}

The solver of Eq. (\ref{pSe}) uses a numerical procedure based
on the conjugate gradient method and it has  been previously used
within a non-self-consistent atomic
superposition scheme.\cite{ATSUP2}
Equation (\ref{pSe}) is solved for the ground state only since
positrons generally thermalize very quickly\cite{Hautojarvi} in a material
and only one positron is present in the material under normal experimental conditions.
In principle, it is also possible to implement positron state calculations into WIEN2k
and other similar codes.\cite{Takenaka,FeCu,rusz,bdj}
On the other hand, the fact that positrons overlap only slightly with core and
other more localized electrons indicates that the representation
of the positron wave function on dense radial meshes inside muffin-tin spheres
is not necessary and a regular 3D mesh
in the unit cell describes properly the positron behavior.
Indeed, the results of detailed numerical tests with varying 3D mesh spacing
show that a spacing of 0.10 -- 0.15 a.u. (for each direction) is usually enough
to obtain numerically accurate results for the positron characteristics.
The exceptions are very light elements such as Li, which
require two or three times denser meshes because of their very small core electron radii.
Thus, our computational scheme employs a real space method for positron calculations
whereas the electronic structure calculations make use of the mixed APW+lo/LAPW basis set.
Our finite difference approach for the positron fully suits to the needs
of the present study and it can be further optimized as suggested
by Sterne \emph{et al.}\cite{sterne99}

In the case of the GGA for positrons, an accurate calculation of the gradient of the electron density is required.
The $\epsilon$ parameter -- when expressed fully in terms of the electron density -- can be written as follows
\begin{equation}
\epsilon = \frac{\pi |\nabla \rho^-|^2}{4 (\rho^-)^2 (3\pi^2 \rho^-)^\frac13} \,,
\end{equation}
which includes the gradient size on the 3D mesh.
In practice, the gradient calculation is done separately for core and valence (including semicore) electrons.
The core electron density is represented on a radial logarithmic mesh inside the muffin-tin spheres.
Therefore, a radial derivative of such densities is first calculated which is then
projected\footnote{The direction of the density gradient is either inward or outward the corresponding nucleus.} as a true gradient on the 3D mesh.
In the case of valence electrons, a 7-point formula\cite{3Dgrad} is used to calculate
the gradient directly on the 3D mesh.
Both gradients are summed up and resulting gradient size is determined.

\subsection{Controlling WIEN2k parameters}  \label{AWP}

The WIEN2k code enables us to control important numerical parameters of electronic structure calculations.
These include muffin-tin radius $R_{mt}$, number of points in the radial muffin-tin mesh, the energy parameters of radial basis functions,\cite{APWlo} the cutoff of the basis set characterized by
the product $R_{mt}K_{max}$ with $K_{max}$ being the maximum size of the $\bm{K}$ vectors in the reciprocal lattice, the cutoff of the lattice harmonics expansion of the electron density and potential in the muffin-tin spheres,
an analogous cutoff for the density and potential plane wave expansion in the interstitial space, \emph{etc.}
As for muffin-tin radius, we use the values recommended by the WIEN2k program (for example, $R_{mt}=2.5$ a.u. for Al) and the numerical values of calculated lifetimes are not very sensitive with respect to $R_{mt}$ unless neighboring muffin-tin spheres are almost touching. Other mentioned WIEN2k parameters also cause very small changes in calculated positron characteristics if their values are close to defaults.
The only exception is the basis set cutoff $R_{mt}K_{max}$,
which needs to be usually somewhat increased to 8 (or 9) from
its default value 7 (especially when the
positron gradient correction is applied)
in order to obtain precise results.

Concerning the $\bm{k}$-point mesh, we kept the product of the number of atoms in the unit cell and the number of $\bm{k}$-points in the whole Brillouin zone constant at a value of $\sim$3000, which appears to be appropriate for positron calculations.
The level of self-consistency expressed in terms of charge convergence\footnote{The charge convergence parameter is defined approximately as the root mean square of the electron density difference in muffin-tin spheres between two subsequent iterations.} was normally 0.00001 or better.

\subsection{Defect calculations}  \label{ADC}

When doing positron defect calculations, the Coulomb potential in the supercell requires a slight adjustment (constant shift) because of different Coulomb potential reference levels in the WIEN2k calculations for the bulk and corresponding supercell.
We evaluated shifts to be $-0.06$, $-0.05$ and $-0.13$ eV for Al, Si and Cu supercells, respectively,
by aligning\footnote{The supercells are constructed in the way that there is an atomic plane at their faces (of \{100\} type). The same applies to the corresponding bulk cells. Positrons tend to have the largest probability density at the minima of the positron potential, which in our cases correspond to the minima of the Coulomb potential acting on positrons. Then, it is plausible to shift the supercells' Coulomb potential in the way that the Coulomb potential minima are aligned at the supercell faces and at the bulk.}
the positron Coulomb potential minima at the bulk and supercell faces.
This potential adjustment affects the calculated lifetime almost negligibly (by 0.1--0.2 ps), but it causes an increase of the positron binding energy to vacancy defects by a non-negligible amount.
Hence, one can use appropriate supercell sizes for accurate positron calculations whereas supercells employed in electronic structure calculations can be much smaller.
In particular, our supercells for Al, Si and Cu positron calculations contained 499, 1727, and 863 atoms, respectively, with the convergence with respect to the supercell size being checked for each material.

For defect calculations, we also note that the positron results appear to be less sensitive to WIEN2k parameters like \RK, and the spacing of the 3D mesh can be somewhat smaller compared to bulk calculations, probably due to smaller overlap of the positron wave function and localized electron ones.

\bibliography{ggam}

  \end{document}